# Structured Optical Fields Reveal Nanoscale Chiral Light-Matter Interactions Governed by Optical Chirality

**Atsushi Kamegaya[1], Shun Hashiyada[1,2,*], and Yoshito Y. Tanaka[1,†]**

[1]Laboratory of Nanosystem Optical Manipulation, Section of Photonics and Optical Science, Research Institute for Electronic Science, Hokkaido University, Nishi 10, Kita 21, Kita-ku, Sapporo, Hokkaido 001-0021, Japan

[2]PRESTO, Japan Science and Technology Agency, 4-1-8 Honcho, Kawaguchi, Saitama 332-0012, Japan

**Corresponding authors**

*shun.hashiyada@es.hokudai.ac.jp

†ytanaka@es.hokudai.ac.jp

## ABSTRACT

Optical chirality has been proposed as the local electromagnetic quantity governing chiral light–matter interactions, yet in conventional circularly polarized fields its magnitude is locked to the electric energy density, obscuring its independent role. Here we create a structured optical field in which optical chirality arises spatially in magnitude and sign while the electric energy density remains nearly uniform. A single chiral nanoparticle exhibits a differential response that follows this spatial variation, whereas no modulation is observed for an achiral nanoparticle, providing direct experimental evidence that optical chirality governs nanoscale chiral light–matter interactions. Measurements of wavelength-dependent optical rotation further provide an experimental estimate of the chiral polarizability, predicting a chiral gradient force of approximately 100 fN and a one-dimensional trapping potential exceeding the thermal energy at room temperature under optimized aqueous trapping conditions.

## TEXT

## I. INTRODUCTION

Optical chirality, originally introduced by Lipkin as a conserved quantity of Maxwell's equations [1] and later identified by Tang and Cohen as the electromagnetic quantity governing differential interactions between chiral matter and electromagnetic fields [2], has become a central concept in chiral photonics. Conventional circular dichroism (CD) spectroscopy probes the differential interaction of chiral matter with left- (LCP) and right-circularly polarized (RCP) plane waves [3]. For such plane waves, however, optical chirality is not independently tunable from the electric energy density: their magnitudes are both proportional to the electric field intensity, while the sign of optical chirality is determined by the polarization handedness. Conventional measurements with circularly polarized light therefore cannot independently establish whether the local chiral response is governed by optical chirality itself or merely accompanies a change in the local electric energy density.

The local nature of optical chirality opens a qualitatively different possibility. Because optical chirality is defined at every point in space, structured optical fields can be designed in which optical chirality and electric energy density have different spatial distributions. Chiral near fields [4,5] and structured interference fields [2,6,7] provide important examples of such field engineering. In particular, Cameron and co-workers proposed an interference field formed by two orthogonally polarized plane waves in which electric energy density remains spatially uniform while optical chirality varies sinusoidally in both magnitude and sign [7]. This field provides a stringent test of optical chirality as an independent local interaction quantity: if the response of a chiral object is governed by optical chirality, it should vary spatially even though electric energy density remains unchanged.

Chirality-engineered fields have also been proposed for chirality-dependent optical forces and enantioselective optical manipulation [8–15]. Early proposals considered enantiomer-dependent optical potentials in standing waves of circularly polarized light [8], and subsequent experiments demonstrated chirality-dependent forces using tightly focused circularly polarized beams [14]. In these configurations, however, gradients of optical chirality coexist with gradients of optical energy density, making the chiral contribution difficult to isolate from conventional achiral optical forces. Brasselet and co-workers later observed chirality-dependent motion of micrometer-scale chiral liquid-crystal droplets in a chirality-modulated field with uniform energy density [13]. Although this demonstrated a chirality-dependent mechanical response to a chirality gradient, particle motion is a consequence of the resulting optical force rather than a direct measurement of the position-dependent chiral response. Direct experimental mapping of such an optical response while the electric energy density remains uniform has therefore remained lacking.

Here, we experimentally realize such a structured optical field and use a single chiral gold nanoparticle [16] as a local probe. The differential response of the chiral nanoparticle follows the local spatial variation of optical chirality, whereas no corresponding modulation is observed for an achiral gold nanosphere. Moreover, changing the modulation period of the structured field produces the corresponding change in the measured nanoparticle response. To our knowledge, these measurements provide the first direct experimental demonstration that nanoscale chiral light-matter interactions are governed by local optical chirality even when the electric energy density remains nearly uniform. We further measure the optical rotatory dispersion (ORD) of the nanoparticles in water to determine their chiral polarizability under conditions relevant to optical trapping. The resulting estimate of the conservative chiral gradient force extends this structured-field approach toward enantioselective optical trapping and separation at the nanoscale.

## II. STRUCTURED OPTICAL FIELD WITH SPATIALLY DECOUPLED CHIRALITY AND ENERGY DENSITY

To spatially decouple optical chirality from electric energy density, we consider the interference of two monochromatic plane waves with orthogonal linear polarizations propagating at symmetric angles $\pm\theta$ in a medium of refractive index $n$, as illustrated in Fig. 1(a). The electric fields of the $p$- and $s$-polarized waves, with vacuum wavelength $\lambda$ and angular frequency $\omega$, are written as $\mathbf{E}_i = (E_0/\sqrt{2})\boldsymbol{\epsilon}_j \exp\{i(\mathbf{k}_j \cdot \mathbf{r} + \delta_j - \omega t)\}$, where $j = p, s$, $\boldsymbol{\epsilon}_p = (\cos\theta, 0, \sin\theta)$, $\boldsymbol{\epsilon}_s = (0, 1, 0)$, $\mathbf{k}_p = k(-\sin\theta, 0, \cos\theta)$, $\mathbf{k}_s = k(\sin\theta, 0, \cos\theta)$, and $k = 2\pi n/\lambda$. The corresponding time-averaged electric energy density $U_e \equiv (\varepsilon_0/4)\|\mathbf{E}\|^2$ and optical chirality density $C \equiv -(\varepsilon_0\omega/2)\mathrm{Im}(\mathbf{E}^* \cdot \mathbf{B})$ are given by

$$U_e(x, \Delta\delta) = \frac{\varepsilon_0}{4} E_0^2 , \tag{1}$$

$$C(x, \Delta\delta) = -k \frac{\varepsilon_0}{2} E_0^2 \cos^2\theta \sin\left\{2\pi\left(\frac{2n\sin\theta}{\lambda}\right)x + \Delta\delta\right\}, \tag{2}$$

where $\Delta\delta = \delta_p - \delta_s$. Equation (1) shows that $U_e$ remains spatially uniform, whereas Eq. (2) shows that $C$ varies sinusoidally in both magnitude and sign, with a modulation period $\Lambda = \lambda/(2n\sin\theta)$. Thus, unlike a circularly polarized plane wave, in which the magnitudes of $C$ and $U_e$ are locked together, the present structured field allows $C$ to vary spatially while $U_e$ remains unchanged. It therefore satisfies $\nabla U_e = 0$, $\nabla C \neq 0$, realizing the spatial decoupling proposed by Cameron and co-workers [7].

Figure 1(b) illustrates the optical system developed to generate this structured field. The input polarization of a He–Ne laser ($\lambda = 632.8\ nm$) was adjusted to provide equal amplitudes of the s- and p-polarized components, and their relative phase $\Delta\delta$ was controlled using a phase retarder. A cylindrical lens focused the beam along one direction at the back focal plane of the objective lens, and a calcite beam displacer separated the s- and p-polarized components into two laterally displaced beams. The two beams were directed to symmetric positions at the back focal plane, producing plane waves incident at angles $\pm\theta$ at the sample plane. For a lateral displacement $d$ of each beam from the optical axis, the incidence angle is determined by $\sin\theta = d/f$, where $f$ is the focal length of the objective lens. The modulation period is therefore $\Lambda = f\lambda/(2nd)$. Thus, for a fixed beam displacement, the spatial period of $C$ can be tuned by changing the focal length of the objective lens without altering the basic field architecture.

For the field characterization shown in Fig. 1(c), the relative phase $\Delta\delta$ was fixed to obtain a static spatial polarization distribution. The transmitted field was magnified using a 60× objective lens (NA = 0.9, $f$ = 3 mm) and a tube lens and recorded at the image plane using an sCMOS camera. A quarter-wave plate and a linear polarizer were inserted into the detection path to separately measure the LCP and RCP components. For the 40× objective lens ($f_{40x}$ = 4.5 mm), the measured LCP and RCP components exhibit complementary stripe patterns with opposite phases, whereas their sum remains nearly uniform. The latter confirms that the spatial polarization modulation is not accompanied by a corresponding intensity modulation. From these images, we calculated the degree of circular polarization, $P_{CP} \equiv (I_{LCP} - I_{RCP})/(I_{LCP} + I_{RCP})$. For the present field geometry, $P_{CP}$ has the same sinusoidal spatial dependence as $C$, although their absolute normalizations differ. The measured $P_{CP}$ therefore directly visualizes the spatial modulation period and phase of the designed optical-chirality distribution. Replacing the 40× objective lens with a 10× objective lens ($f_{10x}$ = 18 mm = 4$f_{40x}$), while keeping the beam displacement fixed, increased the modulation period by approximately a factor of four, consistent with $\Lambda = f\lambda/(2nd)$. These measurements demonstrate the key property of the structured field: optical chirality is spatially modulated while the electric energy density remains nearly uniform.

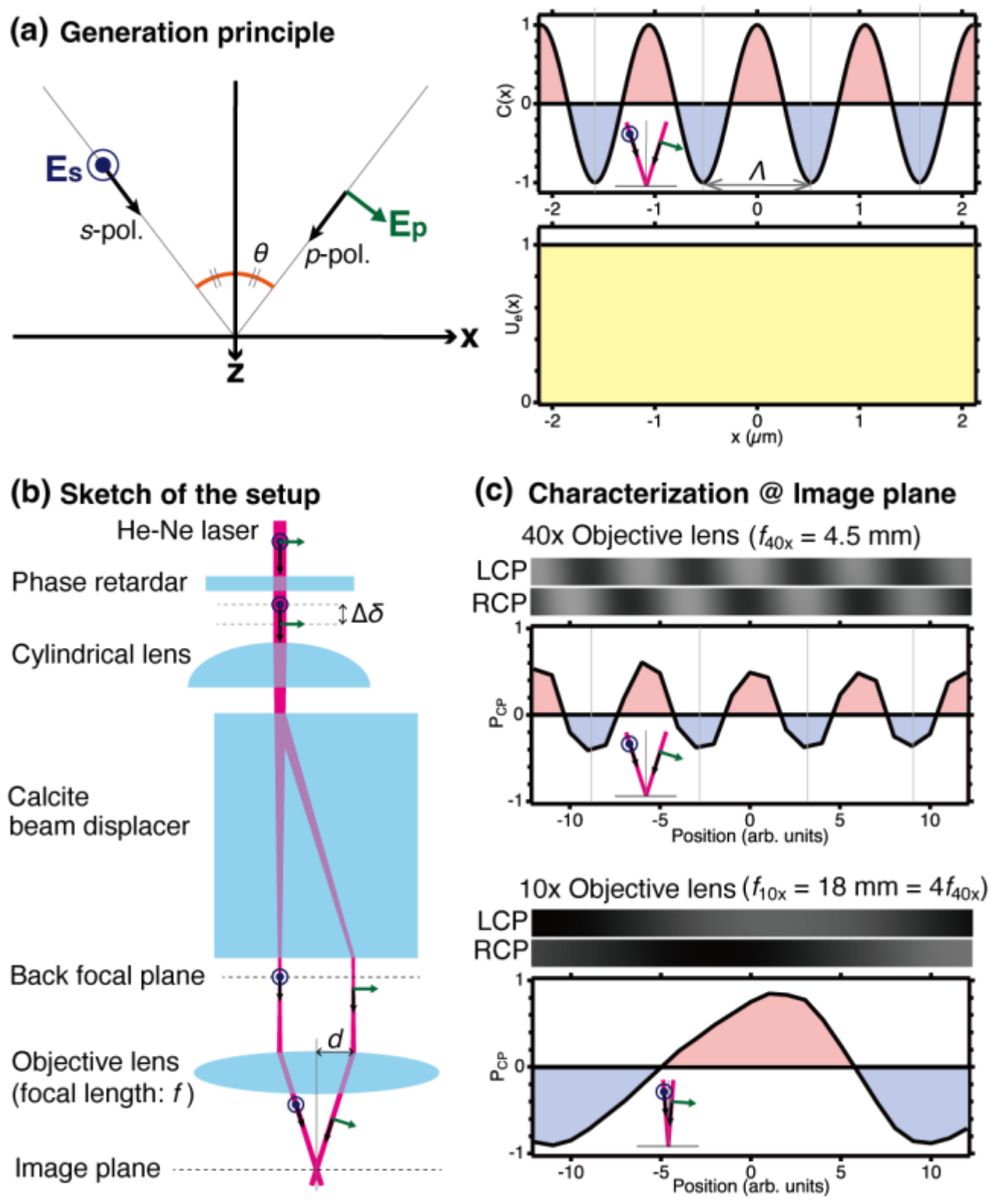

FIG. 1. Generation and characterization of a structured optical field with spatially modulated optical chirality. (a) Principle of the interference field formed by two monochromatic plane waves with orthogonal linear polarizations ($s$ and $p$) incident at symmetric angles $\pm\theta$. The calculated spatial distributions of optical chirality and electric energy density are shown for $\lambda = 632.8$ nm, $\theta = 17.46°$, and $\Delta\delta = -\pi/2$, corresponding to a modulation period of $\Lambda \approx 1.055$ μm. (b) Optical system used to generate the structured field. Equal s- and p-polarized components from a He–Ne laser are phase controlled, spatially separated by a calcite beam displacer, and directed to symmetric positions at the back focal plane of the objective lens. (c) Experimental characterization of the field using polarization-resolved imaging. The left- and right-circularly polarized (LCP and RCP) components and the corresponding degree of circular polarization ($P_{CP}$) are shown for 40× ($f = 4.5$ mm) and 10× ($f = 18$ mm) objective lenses. The complementary LCP and RCP fringes, together with their nearly uniform sum, demonstrate spatial polarization modulation without a corresponding intensity modulation. The approximately fourfold increase in the modulation period agrees with the expected scaling with objective focal length.

## III. SINGLE-NANOPARTICLE PROBE OF LOCAL OPTICAL CHIRALITY

Having established a structured optical field in which the optical chirality $C$ varies spatially while the electric energy density $U_e$ remains nearly uniform, we next test whether the local response of an individual nanoparticle follows $C$ independently of $U_e$. The relative phase between the two interfering beams was switched from $\Delta\delta = -\pi/2$ to $+\pi/2$, such that $C(x, -\pi/2) = -C(x, +\pi/2)$, while $U_e(x, -\pi/2) = U_e(x, +\pi/2)$. This phase modulation reverses the local optical chirality without changing the electric energy density, thereby enabling a differential measurement that isolates chirality-dependent optical responses from achiral background contributions.

To relate the measured differential response to the chiral response of the nanoparticle, we employ the effective chiral constitutive relations

$$\mathbf{p} = \varepsilon_0 \alpha \mathbf{E} + i\frac{\chi}{c\mu_0}\mathbf{B}, \tag{3}$$

$$\mathbf{m} = -i\frac{\chi}{c\mu_0}\mathbf{E} + \frac{\beta}{\mu_0}\mathbf{B}, \tag{4}$$

where $\mathbf{p}$ and $\mathbf{m}$ are the induced electric and magnetic dipole moments, and $\alpha$, $\beta$, and $\chi$ denote the electric, magnetic, and chiral polarizabilities, respectively. Within the effective dipolar description and neglecting the magnetic polarizability $\beta$, the excitation rate of a small chiral object is approximated by [2]

$$A(x, \Delta\delta) \approx 2\omega\left[\mathrm{Im}(\alpha)U_e(x, \Delta\delta) - \mathrm{Im}(\chi)\frac{C(x, \Delta\delta)}{k}\right]. \tag{5}$$

Because the phase modulation reverses $C$ without changing $U_e$, the differential excitation becomes

$$\Delta A(x) = A\left(x, -\frac{\pi}{2}\right) - A\left(x, +\frac{\pi}{2}\right) = -\frac{4\omega}{k}\mathrm{Im}(\chi)C(x). \tag{6}$$

Equation (6) shows that the differential excitation directly maps the spatial distribution of the optical chirality. A chiral nanoparticle with $\mathrm{Im}(\chi) \neq 0$ should therefore exhibit a spatially modulated differential response, whereas an achiral nanoparticle with $\mathrm{Im}(\chi) = 0$ should exhibit no modulation. In the weak-extinction regime, the differential transmission has the same spatial dependence as $\Delta A(x)$, apart from an overall sign and scale factor. The measured differential transmission is therefore expected to follow $C(x)$ only for a chiral nanoparticle.

Figure 2(a) illustrates the measurement scheme. The chiral particles were 432 helicoid III gold nanoparticles prepared and provided by the authors of Ref. [16], with an overall lateral dimension of approximately 150 nm. Achiral gold nanospheres with a diameter of 150 nm were used as reference particles. Measurements were performed in air on individual nanoparticles immobilized on a glass substrate. A substrate-supported nanoparticle was scanned across the structured optical field along the x direction using a piezoelectric stage with a step size of 0.1 μm while the relative phase was periodically switched between $\Delta\delta = -\pi/2$ and $+\pi/2$ at 50 kHz using a photoelastic modulator. The modulation frequency was much higher than the nanoparticle scanning rate, allowing the two optical-chirality states to be compared at effectively the same particle position. The transmitted intensities for $\Delta\delta = -\pi/2$ and $+\pi/2$ were denoted by $I_+$ and $I_-$, respectively. From these signals, we obtained the normalized differential response, $g(x) = 2\,(I_+ - I_-)/(I_+ + I_-)$, and the average transmitted intensity, $I_{ave}(x) = (I_+ + I_-)/2$, simultaneously using lock-in detection.

The experimental results are shown in Fig. 2(b). The average transmitted intensity remained nearly constant for both the achiral gold nanosphere and the chiral nanoparticle, consistent with the nearly uniform $U_e$ independently established in Fig. 1. In contrast, the differential response remained close to zero for the achiral nanoparticle, whereas the chiral nanoparticle exhibited a pronounced sinusoidal modulation. With the 40× objective lens, the measured modulation period was approximately 1.1 μm, in good agreement with the theoretical value of 1.055 μm. Replacing the 40× objective with a 10× objective increased the measured period to approximately 3.9 μm, consistent with the predicted value of 4.22 μm and the fourfold increase expected from $\Lambda = f\lambda/(2nd)$.

These observations provide the direct test enabled by the structured optical field. First, the nearly constant average transmission confirms that the measured spatial modulation is not accompanied by a corresponding variation in electric energy density. Second, the absence of a differential response for the achiral nanosphere rules out a generic response to the structured field. Third, the chiral-nanoparticle response follows the designed sinusoidal variation of $C(x)$ and reproduces the change in its spatial period when the field geometry is modified. Together, these results demonstrate that the position-dependent chiral response persists when optical chirality is spatially decoupled from electric energy density and is governed by the local optical chirality rather than by variations in the field energy. They therefore provide direct experimental evidence that optical chirality functions as an independent local interaction quantity at the nanoscale.

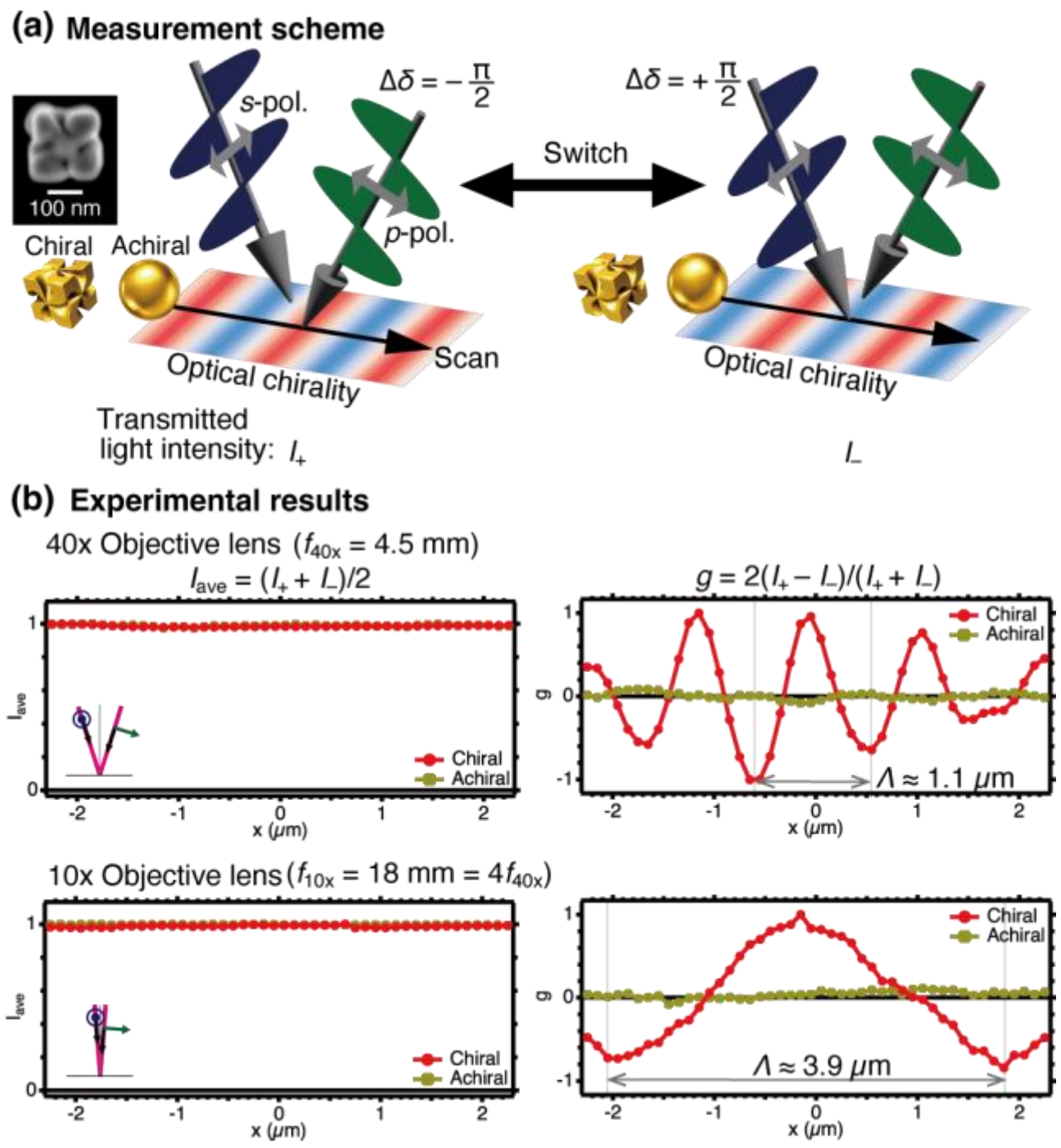

FIG. 2. Single-nanoparticles probing of local optical chirality in the structured optical field. (a) Measurement scheme. A substrate-supported nanoparticle is scanned across the $x$ direction while the relative phase between the two interfering beams is switched at 50 kHz, reversing the local optical chirality without changing the electric energy density. A scanning electron micrograph of the three-dimensional chiral gold nanoparticle used in the measurements is also shown. (b) Normalized average transmitted intensity $I_{ave}$ (left) and normalized differential response $g$ (right) measured for achiral and chiral nanoparticles using 40× ($f$ = 4.5 mm) and 10× ($f$ = 18 mm) objective lenses. The average transmitted intensity remains nearly uniform for both particles. The achiral nanoparticle shows no measurable differential modulation, whereas the chiral nanoparticle exhibits a pronounced periodic response that follows the designed spatial modulation of optical chirality. Changing the objective lens from 40× to 10× increases the measured modulation period from approximately 1.1 μm to 3.9 μm, consistent with the corresponding theoretical period of 1.055 μm and 4.22 μm.

## IV. EXPERIMENTAL ESTIMATION OF THE CHIRAL POLARIZABILITY AND CHIRAL OPTICAL GRADIENT FORCE

Having established in Fig. 2 that the differential response of a single substrate-supported chiral gold nanoparticle follows the local optical chirality, we next assess whether the same chirality-modulated field concept can be extended to nanoscale trapping and separation in water. Because $U_e$ is spatially uniform, the conventional achiral gradient force is suppressed. Whereas the differential excitation in Eq. (6) is governed by $\mathrm{Im}(\chi)$, the conservative chiral gradient force is governed by $\mathrm{Re}(\chi)$ and is given by [9]

$$\mathbf{F}_{chiral}(x) = \frac{\mathrm{Re}(\chi)}{k}\nabla C(x)\,. \tag{7}$$

Here $C(x)$ is given by Eq. (2), and the radiation-pressure contribution is neglected. Evaluating the trapping force therefore requires $\mathrm{Re}(\chi)$, under the aqueous conditions relevant to optical trapping. Because the chiroptical response of a plasmonic nanoparticle depends on its dielectric environment, the polarizability relevant to trapping in water cannot be inferred directly from the substrate-supported measurements in Fig. 2.

We therefore measured the ORD of an aqueous dispersion of the chiral gold nanoparticles used in this study. ORD originates from the difference in the phase accumulated by the LCP and RCP components of linearly polarized light propagating through a chiral medium. The optical rotation angle is given by $\psi_{ORD} = (k_0 l/2)\mathrm{Re}(n_{LCP} - n_{RCP})$, where $k_0 = 2\pi/\lambda$ and $l$ is the optical path length, and $n_{LCP}$ and $n_{RCP}$ are the refractive indices for LCP and RCP light, respectively. For the ORD measurement, linearly polarized light was transmitted through a 1-mm-thick sample cell ($l = 1\ mm$) containing the nanoparticle dispersion, and the transmitted light was analyzed using a linear polarizer. Taking the incident polarization direction as $\phi = 0°$, the transmitted intensities $I_0$, $I_{45}$, $I_{90}$, and $I_{135}$ were measured at analyzer angles of $\phi = 0°$, $45°$, $90°$, and $135°$, respectively. The optical rotation angle was calculated as $\psi_{ORD} =$

$\tan^{-1}[(I_{45} - I_{135})/(I_0 - I_{90})]/2$.

Figure 3(a) shows the measured ORD spectrum, which exhibits a pronounced wavelength-dependent optical rotation. Within the dilute-dispersion approximation and the constitutive convention of Eqs. (3) and (4), the refractive-index difference is related to the particle number density $N$ and the single-particle chiral polarizability $\chi$ through $n_{LCP} - n_{RCP} \approx 2N\chi$. Using the nominal particle number density $N_0 \sim 1.0 \times 10^{15}\, m^{-3}$ [16], the single-particle chiral polarizability was estimated as

$$\mathrm{Re}(\chi) = \frac{\psi_{ORD}}{N_0 k_0 l}. \tag{8}$$

The resulting $\mathrm{Re}(\chi)$ spectrum is shown in Fig. 3(b), together with the corresponding $\mathrm{Im}(\chi)$ spectrum obtained through the Kramers–Kronig relation. The magnitude of $\mathrm{Re}(\chi)$ is approximately $8 \times 10^{-22}\, m^3$ at $\lambda$ = 650 nm. The spectral profile of $\mathrm{Im}(\chi)$ is qualitatively consistent with the characteristic CD response previously reported for 432 helicoid III nanoparticles [16], supporting the assignment of the measured ORD to the chiroptical response of the nanoparticles in water. Because the actual particle number density of the stored dispersion was not independently determined, the absolute magnitude of $\chi$ should be regarded as an order-of-magnitude estimate.

Using the experimentally estimated aqueous $\mathrm{Re}(\chi)$, we evaluated the feasibility of optical trapping in water under conditions optimized independently of those used for the substrate-supported single-particle measurements. The wavelength was chosen as $\lambda$ = 650 nm, where the magnitude of the aqueous $\mathrm{Re}(\chi)$ is larger. From Eqs. (2) and (7), the angular factor entering the chiral gradient force is proportional to $\cos^2\theta\sin\theta$, which is maximized at $\sin\theta = 1/\sqrt{3}$. For water with $n = 1.33$, this condition corresponds to $n\sin\theta = 0.77$. The chiral gradient force $\mathbf{F}_{\mathrm{chiral}}(x)$ and potential $U_{\mathrm{chiral}}(x)$, related by $\mathbf{F}_{\mathrm{chiral}}(x) = -\nabla U_{\mathrm{chiral}}(x)$, were therefore calculated at $\lambda$ = 650 nm in water under the optimized illumination condition ($n\sin\theta = 0.77$), as shown in Fig. 3(c).

For a total optical power density of 5 mW $\mu\mathrm{m}^{-2}$, summed over the two interfering beams, the maximum chiral gradient force reaches approximately 100 fN. The corresponding one-dimensional trapping potential along $x$ has a well depth of approximately $1.3 \times 10^{-20}$ J, exceeding $k_{\mathrm{B}}T$ at $T = 300$ K. The predicted chiral gradient force is therefore sufficiently large to produce chirality-dependent confinement along the modulation direction against thermal fluctuations. Because $\mathbf{F}_{\mathrm{chiral}}$ is proportional to $\mathrm{Re}(\chi)$, opposite enantiomers experience forces in opposite directions and consequently inverted potentials. Their stable positions along the modulation direction are separated by half of the chirality-modulation period ($\Lambda/2 \approx 212$ nm), demonstrating the feasibility of enantioselective lateral confinement and spatial separation at the nanoscale. Together, the substrate-supported local-response measurements and the aqueous ORD-based force estimate establish a route from the direct observation of optical-chirality-governed nanoscale interactions to enantioselective optical manipulation in solution.

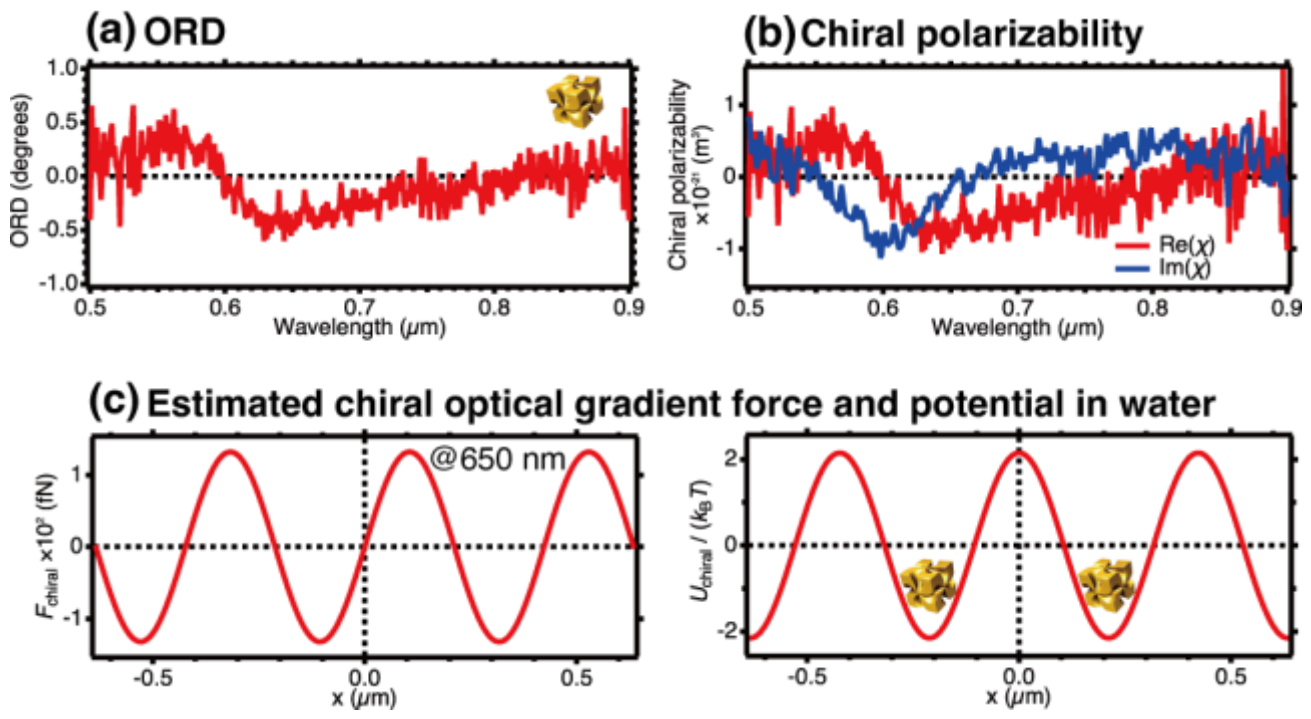

FIG. 3. Experimental estimation of the chiral polarizability and chiral optical gradient force. (a) Optical rotatory dispersion (ORD) spectrum measured for an aqueous dispersion of 432 helicoid III gold nanoparticles using a 1-mm optical path length. (b) Real part of the chiral polarizability, $\mathrm{Re}(\chi)$ (red), estimated from the ORD spectrum using the nominal particle number density $N_0 = 1.0 \times 10^{15}$ m$^{-3}$, together with the corresponding $\mathrm{Im}(\chi)$ (blue) obtained through the Kramers–Kronig relation. (c) Chiral gradient force $\mathbf{F}_{\mathrm{chiral}}$ (left) and one-dimensional trapping potential $U_{\mathrm{chiral}}$ (right) calculated in water at $\lambda$ = 650 nm and $n \sin\theta = 0.77$ for a total optical power density of 5 mW μm$^{-2}$. The potential depth exceeds $k_B T$ at $T$ = 300 K. Opposite enantiomers experience inverted potentials, with stable positions separated by half the chirality-modulation period, $\Lambda/2 \approx 212$ nm.

## V. CONCLUSION

In conclusion, we experimentally realized a structured optical field in which optical chirality varies spatially while the electric energy density remains nearly uniform, and used it to probe local chiral light–matter interactions at the single-nanoparticle level. The differential response of an individual chiral gold nanoparticle followed the designed spatial variation of optical chirality, whereas no corresponding modulation was observed for an achiral gold nanosphere. Moreover, changing the modulation period of the structured field produced the corresponding change in the measured nanoparticle response. These observations provide direct experimental evidence that optical chirality governs the local chiral response of a nanoscale object independently of variations in the electric energy density.

ORD measurements of the nanoparticles in water further provided an experimental estimate of the chiral polarizability relevant to optical trapping in solution. Using the estimated $\mathrm{Re}(\chi)$, we evaluated the conservative chiral gradient force and the corresponding one-dimensional potential. Under optimized aqueous trapping conditions, a total optical power density of 5 mW $\mu m^{-2}$ produces a maximum force of approximately 100 fN and a potential depth exceeding $k_B T$ at room temperature. Although the absolute magnitude depends on the nominal particle number density used to estimate $\chi$, these results support the feasibility of enantioselective optical confinement and spatial separation at the nanoscale.

The present work establishes optical chirality as an experimentally accessible local electromagnetic quantity that governs nanoscale chiral light–matter interactions when spatially decoupled from the electric energy density. More broadly, it demonstrates how structured optical fields can isolate the role of optical chirality and provides a route from fundamental tests of chiral light–matter interactions to enantioselective optical manipulation.

## ACKNOWLEDGEMENTS

We thank Dr. Keiji Sasaki for valuable discussions and Dr. Hyo-Yong Ahn, Dr. Hye-Eun Lee, and Dr. Ki Tae Nam for providing the 432 helicoid III gold nanoparticles used in this study. The authors acknowledge Japan Society for the Promotion of Science (JSPS) KAKENHI (Grant Nos. JP24H00424 and JP22H05132 in Transformative Research Areas (A) “Chiral materials science pioneered by the helicity of light" to Y.Y.T.; JP26K01453 and JP23K04669 to S.H.), and JSPS International Joint Research Program (JRP-LEAD with UKRI) (Grant No. JPJSJRP20241710); Japan Science and Technology Agency (JST) FOREST Program (Grant No. JPMJFR213O to Y.Y.T.) and PRESTO (Grant No. JPMJPR25J6 to S.H.); Ministry of Education, Culture, Sports, Science and Technology (MEXT), Japan, Crossover Alliance to Create the Future with People, Intelligence and Materials (Grant No. 2025Y002 to S.H.).